\documentclass[11pt]{article}
\usepackage{times}
\usepackage{epic,eepic,amssymb,amsmath,amsthm,latexsym}
\usepackage{epsfig}
\usepackage{euler}

\newcommand{\jbox}{\hspace*{\fill} $\Box$}

\newcommand{\dagg}[1]{\ensuremath{ #1^{\mbox{\dag}}}}

\newcommand{\SFT}[1]{\ensuremath{ \mbox{SFT}(#1)}}

\newcommand{\K}{\ensuremath{\mathbb{K}}}

\newcommand{\boolepp}{\ensuremath{\mathbb B}}

\newcommand{\Z}{\ensuremath{\mathbb Z}}

\newcommand{\B}{\boolepp}
\newcommand{\Q}{\ensuremath{\mathbb Q}}

\newcommand{\enth}[1]{\ensuremath{#1^{\mbox{\scriptsize th}}}}

\newtheorem{theorem}{Theorem}[section]
\newtheorem{corollary}[theorem]{Corollary}
\newtheorem{lemma}[theorem]{Lemma}

\newtheorem{proposition}[theorem]{Proposition}
\newtheorem{definition}[theorem]{Definition}

\newenvironment{listedense}{\begin{list}{$\bullet$}{\setlength{\parsep}{0pt}
        \setlength{\parskip}{0pt}
        \setlength{\topsep} {0pt} \setlength{\itemsep} {0pt}
        \setlength{\labelsep}{1em}}}{\end{list}}

\newenvironment{indense}{\begin{list}{$\hspace{1em}$}{\setlength{\parsep}{0pt}
        \setlength{\parskip}{0pt}
        \setlength{\topsep} {0pt} \setlength{\itemsep} {0pt}
        \setlength{\labelsep}{1em}}}{\end{list}}

\newcounter{zahl}
\newenvironment{comptedense}{\begin{list}{$\arabic{zahl}.$}{\usecounter{zahl}
        \setlength{\parsep}{0pt}
        \setlength{\parskip}{0pt}
        \setlength{\topsep} {0pt} \setlength{\itemsep} {0pt}
        \setlength{\labelsep}{1em}}}{\end{list}}

\newenvironment{latindense}{\begin{list}{$\roman{zahl}.$}{\usecounter{zahl}
        \setlength{\parsep}{0pt}
        \setlength{\parskip}{0pt}
        \setlength{\topsep} {0pt} \setlength{\itemsep} {0pt}
        \setlength{\labelsep}{1em}}}{\end{list}}

\newcommand{\bool}{{\mathbb B}}

\newcommand{\C}{{\mathbb C}}

\newcommand{\ket}[1]{\ensuremath{|#1\rangle}}

\newcommand{\mat}[2]{{\mathbb M}_{#1}^{#2}}

\newcommand{\transpose}[1]{#1^{\mathsf{T}}}%
\newcommand{\Transpose}[1]{\left( {#1} \right)^{\mathsf{T}}}%

\newcommand{\ten}[2]{{\mathbb T}_{#1}^{#2}}

\newcommand{\val}[1]{\textnormal{val}_{#1}}

\newcommand{\PSPACE}{\ensuremath{\textnormal{PSPACE}}}

\newcommand{\MODP}[1]{\ensuremath{{\textnormal{MOD}_{#1}\textnormal{-P}}}}
\newcommand{\GapP}{\ensuremath{{\textit{Gap}\textnormal{P}}}}

\newcommand{\nP}{\ensuremath{\#\textnormal{P}}}
\newcommand{\Poly}{\ensuremath{\textnormal{P}}}
\newcommand{\NP}{\ensuremath{\textnormal{NP}}}

\newcommand{\NCone}{\ensuremath{\textnormal{NC}^1}}
\newcommand{\ACzero}{\ensuremath{\textnormal{AC}^0}}
\newcommand{\ACCzero}{\ensuremath{\textnormal{ACC}^0}}

\newcommand{\PP}{\ensuremath{\textnormal{PP}}}
\newcommand{\BPP}{\ensuremath{\textnormal{BPP}}}
\newcommand{\BQP}{\ensuremath{\textnormal{BQP}}}
\newcommand{\NQP}{\ensuremath{\textnormal{NQP}}}

\newcommand{\coCP}{\ensuremath{\textnormal{coC}_=\textnormal{P}}}

\newcommand{\resp}{re\-spec\-tive\-ly}

\date{}

\begin{document}

\title{\bf A common algebraic description for probabilistic
and quantum computations\thanks{Supported by the Qu{\'e}bec FCAR,
by the NSERC of Canada, and by Deutsche Forschungsgemeinschaft.
}}


\author{Martin Beaudry\thanks{Corresponding author.}\\
  Universit{\'e} de Sherbrooke
  \and 
Jos\'e M. Fernandez\\
    Universit{\'e} de Montr{\'e}al
  \and 
  Markus Holzer\thanks{Part of the work was done while the author was
    at D{\'e}partement d'I.R.O., Universit{\'e} de Montr{\'e}al.
\newline 
Beaudry:
  D{\'e}partement de math{\'e}matiques et d'informatique,
  Universit{\'e} de Sherbrooke, 2500 boul. Universit{\'e},
  Sherbrooke, Qu{\'e}bec, J1K 2R1 Canada.
  email: \texttt{beaudry@dmi.usherb.ca}
\newline
Fernandez:
  D{\'e}partement d'I.R.O.
    Universit{\'e} de Montr{\'e}al,
    C.P.~6128, succ.\ Centre-Ville,
    Montr{\'e}al, Qu{\'e}bec, H3C\,3J7 Canada.
  email: \texttt{fernandz@iro.umontreal.ca}
\newline
Holzer:
  Institut f\"ur Informatik, Technische Universit\"at M\"unchen,
  Arcisstra\ss e 21, D-80290 M\"unchen, Germany.
  email: \texttt{holzer@informatik.tu-muenchen.de}
}\\
  Technische Universit\"at M\"unchen}

\maketitle


\begin{abstract}
We study the computational complexity of
the problem SFT (\emph{Sum-free Formula partial Trace})~:
given a tensor formula $F$ over a subsemiring of 
the complex field $(\C,+,\cdot)$ plus a positive
integer $k$, under the restrictions that
all inputs are column vectors of $L_2$-norm 1 and
norm-preserving square matrices,
and that the output matrix is a column vector,
decide whether the $\enth{k}$ partial trace of $F\dagg{F}$ is superior to $1/2$.
The $\enth{k}$ partial trace of a matrix is the sum of its
lowermost $k$ diagonal elements.
We also consider the promise version of this problem, where 
the $1/2$ threshold is an isolated cutpoint.
We show how to encode a quantum or reversible gate array into a
tensor formula which satisfies the above conditions, and vice-versa;
we use this to show that the promise version of SFT is complete for 
the class $\BPP$ for formulas over the semiring $(\Q^+,+,\cdot)$
of the positive rational numbers, for
$\BQP$ in the case of formulas defined over the field $(\Q,+,\cdot)$, and for
$\Poly$ in the case of formulas defined over the Boolean semiring, all
under logspace-uniform reducibility.
This suggests that the difference
between probabilistic and quantum polynomial-time computers
may ultimately lie in the possibility, in the latter case, of having
destructive interference between computations occuring in parallel.
\end{abstract}

\setlength {\baselineskip} {16pt}


\section{Introduction}
\label{sec:introduction}

The ``algebraic approach'' in the theory of computational complexity
consists in characterizing complexity classes within unified
frameworks built around a computational model or problem involving
an algebraic structure (usually finite or finitely generated) as
the main parameter. 
In this way, various complexity classes are seen to share the same definition, 
up to the choice of the underlying algebra.
Successful examples of this approach include the description
of $\NCone$ and its subclasses $\ACzero$ and $\ACCzero$ in terms of polynomial-size
programs over finite monoids \cite{bath88}, and analogous results for
\PSPACE, the polynomial hierarchy and the polytime mod-counting classes,
through the use of polytime leaf languages \cite{helascvowa93}. 
A more recent example is the complexity of problems whose input is a
tensor formula, i.e. a fully parenthetized expression where the inputs are
matrices (given in full) over some finitely generated algebra and the allowed
operations are matrix addition, multiplication, and tensor product
(also known as outer, or direct, or Kronecker product).
Depending on the semiring over which the formula
is defined, the problem of deciding whether the output matrix contains
at least one nonzero entry is complete for \NP\ (Boolean semiring) and
$\MODP{q}$ (modulo semiring $\Z_q$) \cite{dahomc00}.
Other common-sense computational problems on tensor formulas were analyzed
in \cite{dahomc00,beho01}.
\newline
Tensor formulas are a compact way of specifying very large matrices.
As such, they immediately find a potential application in the description
and the behavior of circuits, be they classical Boolean, arithmetic
(tensor formulas over the appropriate semiring)
or quantum (formulas over the complex field, or an adequately chosen
subsemiring thereof).
In this paper, we formalize and confirm this intuition, in that we define
a meaningful computational problem over tensor formulas which enables us
to capture the significant complexity classes \Poly, \BPP, and \BQP. 
Looking at variants of the problem enables us to capture further complexity classes;
a table in the last section summarizes our results.
\newline
Apart from offering a first application of the algebraic approach to
quantum computing, our paper reasserts the point made by Fortnow \cite{for00},
that for the classes \BPP\ and \BQP, the jump from classical to quantum 
polynomial-time computation consists in allowing negative
matrix entries for the evolution operators,
which means the possibility of having destructive interference
between different computations done in parallel.


\section{Background on circuits and complexity}
\label{sec:complexity}

We use standard notions and  notations from computational complexity,
see for example \cite{badiga95,weg87}.
In particular we assume
that the reader is familiar with the
deterministic and probabilistic Turing machine models,
with the usual notion of a Boolean circuit,
%
%
and with logspace many-one reducibility: a set $K$ is \emph{logspace
time many-one reducible} to a set $L$ if there is a 
logspace computable mapping~$f$
such that for all~$x$, $x \in K$ iff $f(x) \in L$.
\newline
To handle the three types of computation discussed in this paper 
(deterministic, probabilistic and quantum), we use 
\emph{gate arrays} as a common setting.
From now on, we reserve the word \emph{circuit} to the traditional idea
of an acyclic network with a unique output bit,
and we use \emph{gate array} to describe
those computational networks which satisfy the following definition.

\begin{definition}
\label{def:gatearray}
Let $n,d \ge 1$. A \emph{ width $n$, $d$-leveled gate array} 
is a $n \times d$
array where each line is called a \emph{wire}
and each column a  \emph{level}.
The \emph{size} of a gate array is the number $nd$.
A \emph{gate} is a set of array entries from the same level
(corresponding to the wires involved in the gate's operation)
together with a square matrix which describes its action.
Gates on a given level act on
disjoint sets of entries from this level.
Let the levels be numbered $1$ to $d$ from left to right.
Each wire carries a bit from a level to the next in the
left-to-right direction; the value entering
column $1$ from the left is called an \emph{input} the value
exiting level $d$ to the right is an \emph{output}.
\end{definition}

A gate of~$k$ binary inputs operates on the set of $k$-bit
vectors by mapping each of the $2^k$ possible combinations of 
input values to a combination of output values.
The extra constraint, that all gates act on neighboring
wires, can be enforced on an arbitrary array at the cost of 
inserting a quadratic number of extra levels with ``swap'' gates,
which interchange the values carried by two adjacent wires.
\newline
Gate arrays are used in particular to describe
reversible   classical computations.
A computation is \emph{reversible} iff knowledge
of its output is sufficient to
be able to deterministically reconstruct the input.
It has been shown that for any polynomial-time deterministic algorithm
there exists an equivalent polynomial-time reversible algorithm;
in other words, from every polynomial-size Boolean circuit
an equivalent reversible gate array \cite{frto82}\ can be constructed, by
\begin{listedense}
\item  modifying the circuit so that
the numbers of input and output bits are equal;
\item replacing the usual one-output gates with reversible gates;
\item making sure that an especially identified ``decision'' bit
takes value $1$ at the output level iff the original circuit's output is $1$.
\end{listedense}
From the description of the original circuit, 
its equivalent reversible gate array can be constructed in
deterministic logspace;
circuit size and depth are increased only by a polynomial factor;
usually, a polynomial number of extra input bits initialized at $0$,
called \emph{ancillary bits}, also has to be added in the process.
It has been shown that this array can be built solely
with the one- and two-bit reversible operations, plus
either one of the ``Toffoli'' ($\Theta$) or ``Fredkin'' ($\Phi$) gates,
where
$$
\Theta=\left[\begin{matrix}
    1 & 0 & 0 & 0 & 0 & 0 & 0 & 0 \\
    0 & 1 & 0 & 0 & 0 & 0 & 0 & 0 \\
    0 & 0 & 1 & 0 & 0 & 0 & 0 & 0 \\
    0 & 0 & 0 & 1 & 0 & 0 & 0 & 0 \\
    0 & 0 & 0 & 0 & 1 & 0 & 0 & 0 \\
    0 & 0 & 0 & 0 & 0 & 1 & 0 & 0 \\
    0 & 0 & 0 & 0 & 0 & 0 & 0 & 1 \\
    0 & 0 & 0 & 0 & 0 & 0 & 1 & 0 
    \end{matrix}\right]
\quad \text{and} \quad
\Phi=\left[\begin{matrix}
    1 & 0 & 0 & 0 & 0 & 0 & 0 & 0 \\
    0 & 1 & 0 & 0 & 0 & 0 & 0 & 0 \\
    0 & 0 & 1 & 0 & 0 & 0 & 0 & 0 \\
    0 & 0 & 0 & 1 & 0 & 0 & 0 & 0 \\
    0 & 0 & 0 & 0 & 1 & 0 & 0 & 0 \\
    0 & 0 & 0 & 0 & 0 & 0 & 1 & 0 \\
    0 & 0 & 0 & 0 & 0 & 1 & 0 & 0 \\
    0 & 0 & 0 & 0 & 0 & 0 & 0 & 1 
    \end{matrix}\right] ;
$$
here the top left position corresponds to bit values $000$ and the
bottom right to $111$.
\newline
Standard techniques can therefore
be used in sequence to transform the description
of a polytime deterministic Turing machine and its input $x$ into an instance
of the Circuit Value Problem with constant inputs (where $x$ is hardwired)
\cite{lad75},
then to turn this circuit into a reversible gate array, in order
to give the following definition for the class $\Poly$.
(Alternatively, one can start from the definition of
\Poly\ as the class of those languages decided by logspace-uniform
families of polynomial-size Boolean circuits.)

\begin{definition} 
\label{def:polytime}
\Poly\ is the class of those languages $L\subset \Sigma^*$ for which 
there exist a logspace-computable function  which,
given an input $x \in \Sigma^*$, computes the encoding of a
reversible gate array $C(x)$ with constant inputs, whose decision bit takes
value $1$  at the output level iff $x \in L$.
\end{definition}

An encoding for $C(x)$ is suitable for this definition if
it consists of a reasonable description of the array's inputs,
wiring and gates; the latter can wlog be restricted to have
constant fan-in/fan-out, so that the action of each gate can be specified
with a constant-size Boolean matrix.
\newline
Complexity classes for polynomial-time probabilistic computation
are usually defined in terms of a polytime Turing
machine which picks a random bit at every step of its computation, and
otherwise acts deterministically (see e.g. \cite{badiga95}).
An equivalent circuit is built from this Turing machine and its input,
in which an appropriate number of random bits are fed in
alongside the (constant) input bits;
whether the input belongs to $L$ is verified by counting
those combinations of random bits for which the
output bit takes value $1$. All random bit combinations have equal length and
are equally likely.

\begin{definition} 
\label{def:PPandBPP}
\PP\ is the class of those languages $L\subset \Sigma^*$ for which 
there exist a logspace-computable function  which,
given an input $x \in \Sigma^*$, yields the encoding of
a reversible gate array $C(x)$ with a combination 
of constant and random inputs,
such that $x\in L$ iff $f_C(x) > \frac{1}{2} $
and $x\not\in L$ iff $f_C(x) < \frac{1}{2} $,
where $f_C(x)$ denotes the 
probability that $C(x)$'s decision bit takes
value $1$  at the output level.
\newline
\BPP\ is defined with the extra condition that there exists
a parameter $\varepsilon$, $0< \varepsilon
< \frac{1}{2} $, 
such that
$x\in L$ iff $f_C(x) > \frac{1}{2} +\varepsilon$
value $1$  at the output level.
\newline
The class \NP\ can be similarly defined, with the condition that 
$x\in L$ iff $f_C(x) > 0$.
\end{definition}

The definition of \BPP\ includes the implicit constraint, that
the proportion of accepting computations can never 
fall inside the interval  $[\frac{1}{2}-\varepsilon,\frac{1}{2}+\varepsilon]$;
in other words, $\frac{1}{2}$ is an isolated cutpoint.
Note that both \PP\ and \BPP\ can be
redefined with a cutpoint other than $\frac{1}{2}$.\\

Polynomial-time quantum computation was defined originally in terms of
quantum Turing machines \cite{de85}: the data handled by this machine
(\emph{qubits})
are formally represented as a vector whose complex components give the
distribution of amplitudes for the probability that the qubits
be in a certain combination of values;
each transition of the machine acts 
as a unitary transformation on this vector. 
\newline
It was later shown \cite{yao93}\ that a
quantum Turing machine and its input can be encoded in deterministic
polynomial time into an array of quantum gates, if one is
allowed a small probability of error.
Each wire in a quantum gate array represents a path of
a single {qubit} (in time or space, forward from
left to right), and is described by a state in a two dimensional
Hilbert space with basis~$\ket{0}$ and~$\ket{1}$. Just as classical
bit strings can represent the discrete states of arbitrary finite
dimensionality, so a string of~$n$ qubits can be used to represent
quantum states in any Hilbert space of dimensionality up to~$2^n$.
The action of a gate of~$k$ inputs is a
unitary operation of the group $U(2^k)$, i.e., a generalized rotation
in a Hilbert space of dimension~$2^k$.
It has been shown that a small set of one- and two-qubit gates suffices
to build quantum arrays,  
in that any $n$-qubit gate can be simulated by a subarray consisting of
two-qubit gates, and
the number thereof is at most an exponential in $n$
(see for example \cite{babecletal95,di95,slwe95,ll95}).
As two-qubit gates it suffices to take the
\emph{controlled-not} $N$. Because
of its usefulness we also mention the two-qubit ``swap'' gate $T$.
$$
N=\left[\begin{matrix}
      1 & 0 & 0 & 0\\
      0 & 1 & 0 & 0\\
      0 & 0 & 0 & 1\\
      0 & 0 & 1 & 0
  \end{matrix}\right], \quad
T=\left[\begin{matrix}
        1 & 0 & 0 & 0\\
        0 & 0 & 1 & 0\\
        0 & 1 & 0 & 0\\
        0 & 0 & 0 & 1
    \end{matrix}\right] .
$$

The vector of qubits received as input by a quantum gate array can be
regarded as a linear combination of \emph{pure states}.  There is a
\emph{measurement} done on the array's output, which consists in
projecting the output vector onto a subspace, usually defined by
setting a chosen subset of the qubits to $\ket{1}$ (``accepting
subspace'').  If the qubits are numbered $1$ to $n$, then a $k$-qubit
accepting subset can be chosen to be qubits $1$ to $k$, at the cost of
inserting a quadratic number of extra swap gates.  For the sake of
simplicity, we can assume that the final output state will be such
that all qubits other than the decision qubit have value $\ket{0}$.
This is without loss of generality, as it will be possible to
``uncompute'' the circuit while keeping the value of the decision bit.
Thus, the accepting subspace has dimension $1$, and contains only one
base vector, and similarly for the rejecting subspace.

\begin{definition} 
\label{def:BQP}
\BQP\ is the class of those languages $L\subset \Sigma^*$ for which 
there exist
a logspace-computable function  which,
given an input $x \in \Sigma^*$, yields the encoding of a 
quantum gate array $C(x)$ with constant inputs,
and a parameter $\varepsilon$, $0< \varepsilon
< \frac{1}{2} $, 
such that
$x\in L$ iff $f_C(x) > \frac{1}{2} +\varepsilon$
and $x\not\in L$ iff $f_C(x) < \frac{1}{2} -\varepsilon$,
where $f_C(x)$ denotes the 
probability that the qubits of $C(x)$ be projected onto 
the accepting subspace  at the output level.
\end{definition}

The remark on parameter $\varepsilon$ 
made after the definition of $\BPP$ also holds here.
The definition of $\BQP$ still holds if
we restrict the gates to implement unitary operators with entries
taken in a small set of rationals \cite{addehu97},
and to determine acceptance or rejection by the
value of a single qubit \cite{bebebrva97}.
\newline

The same definition, with unitary operators and input vectors having
rational entries and without the condition that $\frac{1}{2}$ be an
isolated cutpoint, yields a ``quantum'' version of the (classical)
class \PP.  However, this ``new'' class is in fact no different than
\PP\ itself, as can be shown by a simple counting complexity theory.

For any language $L$ in this class, there exists a quantum circuit
that accepts it, for which we can define the non-negative functions
$f(x)$ and $g(x)$, as the sum of all the positive and negative
contributions, respectively, to the total amplitude for the accepting
configuration on a given input $x$.  The amplitude of this unique
accepting configuration is $f(x)-g(x)$.  Similarly, define $f'(x)$ and
$g'(x)$ for the rejecting configuration, with the corresponding
rejecting amplitude being $f'(x)-g'(x)$.  It is easy to see that $f$,
$g$, $f'$, and $g'$ are all \nP\ functions.  The difference between
the probability of accepting and rejecting of this circuit is thus
$$
(f-g)^2 - (f'-g')^2 = f^2 + g^2 + 2f'g' - (f'^2 + g'^2 + 2fg)
$$
which is a \GapP\ function, since \nP\ is closed under (finite) sum
and product.  This function will be positive if and only $x$ is in
$L$, which is another way of characterizing languages in the class
\PP\ \cite{for97}.

On the other hand, the languages defined with quantum gate arrays
where unitary operators have rational entries and such $x\in L$ iff
$f_C(x) > 0$ form the complexity class \NQP, the quantum analogue to
\NP, which coincides with the (classical) class \coCP\ 
\cite{fegrhopr99}.


\section{Tensor Algebra}
\label{sec:algebra}

A \emph{semiring} is a tuple $(\K ,+,\cdot)$ with
$\{0,1\}\subseteq\K $ and binary operations $+,\cdot:
\K \times\K \rightarrow\K $ (sum and product),
such that $(\K ,+,0)$ is a commutative monoid,
$(\K ,\cdot,1)$ is a monoid, multiplication distributes over
sum, and $0\cdot a=a\cdot 0=0$ for every~$a$ in~$\K $ (see,
e.g.,~\cite{kusa86}). 
A semiring is a \emph{ring} if and only if $(S,+,0)$ is a group.
In this paper we consider
the following semirings: the Booleans $(\bool,\vee,\wedge)$, 
the field of rational numbers $(\Q,+,\cdot)$,
the semiring $(\Q^+,+,\cdot)$ of positive rational numbers, and 
the field of complex numbers $(\C,+,\cdot)$.

Let $\mat{\K }{}$ denote the set of all \emph{matrices} over
$\K $, and define
$\mat{\K }{k,{\ell}}\subseteq\mat{\K }{}$ to be the
set of all \emph{matrices of order $k\times\ell$}.
Let~$[k]$ denote the set $\{1,2,\ldots,k\}$;
for a matrix~$A$ in~$\mat{\K }{k,\ell}$ and
$(i,j) \in [k]\times[\ell]$, the
$\enth{(i,j)}$ entry of $A$ is denoted by~$a_{i,j}$ or $(A)_{i,j}$.
Addition and multiplication of
matrices in $\mat{\K }{}$ are defined in the usual way.
Additionally we consider the \emph{tensor product}
$\otimes:\mat{\K }{}\times\mat{\K }{}\rightarrow\mat{\K }{}$
of matrices, also known as Kronecker product, outer product, or direct
product, which is defined as follows: for
$A\in\mat{\K }{k,{\ell}}$ and $B\in\mat{\K }{m,n}$ let
$A\otimes B\in\mat{\K }{km,\ell n}$ be
$$
A\otimes B:=\left[
  \begin{array}{ccc}
    a_{1,1}\cdot B&\ldots&a_{1,{\ell}}\cdot B\\
    \vdots & \ddots & \vdots\\
    a_{k,1}\cdot B&\ldots&a_{k,{\ell}}\cdot B
  \end{array}
\right].
$$
Hence $(A\otimes B)_{i,j}=(A)_{q,r}\cdot (B)_{s,t}$ where $i=k\cdot
(q-1)+s$ and $j=\ell\cdot (r-1)+t$.

The following notation is used: let~$I_n$ be the order~$n$
identity matrix, $e_i^n$ the $n \times 1$ column vector whose
$\enth{i}$ entry has value $1$ and the others $0$.
and let~$A^{\otimes n}$ stand for the $n$-fold
iteration $A\otimes A\otimes \cdots \otimes A$.

\emph{Stride permutations}, which play a
crucial role in the implementation of efficient parallel programs for
block recursive algorithms such as the fast Fourier transform (FFT)
and Batcher's bitonic sort (see \cite{toanlu97})
will be useful in our proofs.  
The \emph{$mn$-point stride~$n$
permutation}~$P_n^{mn}\in\mat{\K }{mn,mn}$ is defined as
$$P_n^{mn}{e_i^m\otimes e_j^n}={e_j^n\otimes e_i^m},$$
where $e_i^m\in\mat{\K }{m,1}$ and
$e_j^n\in\mat{\K }{n,1}$. In other words, the
matrix~$P_n^{mn}$ permutes the elements of a vector of length~$mn$
with stride distance~$n$.
We will
make use of the following identities on stride permutations.

\begin{proposition}
\label{prp:basic-stride}
  The following holds for all $\ell,m,n$:
  \begin{comptedense}
  \item $\left(P_n^{mn}\right)^{-1} = P_m^{mn}$;
  \item $ P_{mn}^{\ell mn} = P_m^{\ell mn} \cdot P_n^{\ell mn} $; 
  \item $ P_{n}^{\ell mn} = \left(P_n^{\ell n} \otimes I_m\right)
\cdot \left(I_\ell \otimes P_n^{mn}\right) $. \qed
  \end{comptedense}
\end{proposition}



\subsection{Tensor formulas}

\begin{definition}
\label{def:formula}
The \emph{tensor formulas} over a semiring~$\K $ and their
\emph{order} are recursively defined as follows.
\begin{comptedense}
  \item Every matrix~$F$ from $\mat{\K }{k,\ell}$ with entries
    from~$\K $ is a (\emph{atomic}) tensor formula of order
    $k\times\ell$. 
  \item If~$F$ and~$G$ are tensor formulas of order $k\times\ell$ and
    $m\times n$, \resp, then 
    \begin{indense}
    \item $(F+G)$ is a tensor formula of order is $k\times\ell$ if
      $k=m$ and $\ell=n$;
    \item $(F\cdot G)$ is a tensor formula of order $k\times n$ if
      $\ell=m$;
    \item $(F\otimes G)$ is a tensor formula of order $km\times \ell
      n$.
    \end{indense}
  \item Nothing else is a tensor formula.
\end{comptedense}
We say that a tensor formula $F$ is \emph{sum-free}
whenever none of $F$ and its subformulas has the form $G+H$.
Let~$\ten{\K }{}$ denote the set of all tensor formulas
over~$\K $, and define
$\ten{\K }{k,\ell}\subseteq\ten{\K }{}$ to be the
set of all tensor formulas of order $k\times\ell$.
\end{definition}

In this paper we only consider semiring elements whose value can
be given with a standard encoding over some finite set $\mathcal{G}$.
Input matrices can therefore be
string-encoded using list notation such as ``$[[001][101]]$.'' 
Nonatomic tensor formula can be encoded over the alphabet
$\Sigma= \mathcal{G}\cup\{[,],(,),\cdot,+,\otimes\}$.
Strings over~$\Sigma$
which do not encode valid formula are deemed to represent the trivial
tensor formula~$0$ of order $1\times 1$. 
\newline
The \emph{size} of a tensor formula $F$ is $1$ if $F$ is atomic,
otherwise $F = G \circ H$ for $\circ \in \{+,\cdot,\otimes\}$
and the size of $F$ is $1$ plus the sizes of 
$G$ and $H$. 
The \emph{diameter} of tensor formula $F$, denoted by $|F|$, is
$\max\{k,\ell\}$  if $F$ is atomic of order $k\times\ell$;
otherwise we have that $F = G \circ H$ is of order
$k\times\ell$, and $|F| = \max\{k,\ell,|G|,|H|\}$.
\newline
It will sometimes be convenient to speak of a tensor formula in
graph-theoretical terms: in this context, a tensor formula is
a binary tree whose edges are directed toward the root (``output
node''), whose leaves (``input nodes'')
are labelled with atomic formulas 
and each of whose interior nodes is labelled with an operation from
the set $\{+,\cdot,\otimes\}$. The depth of a tensor formula is the 
maximum root-leaf distance.

\begin{definition}\label{def:val} 
  For each semiring~$\K $ and each~$k$ and each~$\ell$ we
  define
  $\val{\K }^{k,\ell}:\ten{\K }{k,\ell}\rightarrow
  \mat{\K }{k,\ell}$, that is, we associate with node~$f$ of
  order $k\times\ell$ of a tensor formula~$F$ its $k\times\ell$
  matrix ``value,'' which is defined as follows:
  \begin{comptedense}
    \item $\val{\K }^{k,\ell}(f)=F$ if~$f$ is an input node
      labeled with~$F$,
    \item
      $\val{\K }^{k,\ell}(f)=\val{\K }^{k,\ell}(g)+
      \val{\K }^{k,\ell}(h)$ if $f=(g+h)$,
    \item $\val{\K }^{k,\ell}(f)=
      \val{\K }^{k,m}(g)\cdot \val{\K }^{m,\ell}(h)$
      if $f=(g\cdot h)$ , and
    \item $\val{\K }^{k,\ell}(f)=
      \val{\K }^{k/m,\ell/n}(g)\otimes
      \val{\K }^{m,n}(h)$ if $f=(g\otimes h)$ .
    \item For completeness, recall that $\val{\K }^{k,\ell}(f)= 0$
      whenever the formula is not valid.
  \end{comptedense}
The value $\val{\K }^{k,\ell}(F)$ of a tensor formula~$F$ of
order $k\times\ell$ is defined to the value of the unique output node.
%
%
%
\end{definition}

\subsection{The sum-free partial trace problem} 
\label{subsec:sftp}

A column vector $v$ with complex coefficients is a \emph{unit}
vector iff its $L_2$-norm is $1$, that is, iff ${v}^\dag v= 1$.
In this paper, we work on probabilistic and quantum computations
where the probability amplitudes are encoded in unit column vectors,
and the foremost requirement on the computing model is that the
inner product (hence also the $L_2$ norm)
be preserved at each step of a computation.
The action of each such step on the various combinations of values
transported by the wires is described with a square matrix; 
our requirement is equivalent to asking
that each matrix preserves the inner product (\emph{unitary matrices}).
\newline
A square matrix $M$ over the complex numbers is \emph{unitary}
iff ${M}^\dag = M^{-1}$.
For a matrix $M$ over the real numbers, this translates into
$\transpose{M} = M^{-1}$; which means that $M$ is \emph{orthogonal}.
It is an easily verified fact that 
an orthogonal matrix contains only nonnegative entries
if, and only if, it is a permutation matrix (i.e., exactly
one entry per line and column is $1$ and all others are $0$).
\newline
In the sequel, whenever we deal simultaneously with the cases where
matrices with real or complex coefficients, we use the notations and
vocabulary from the real case alone, in order to make the text easier
to read.
\newline
The \emph{trace} of a square matrix is the sum of its
diagonal elements~; for $k>0$, 
its \emph{$\enth{k}$ partial trace} is the sum of its
last $k$ diagonal elements, counting upwards from the lower right corner.
For completeness, if $k$ exceeds the diameter of the matrix, then
the {$\enth{k}$ partial trace} coincides with the usual trace.

\begin{definition}
\label{def:sum-free}
A sum-free tensor formula is \emph{OSL}
if and only if it satisfies the conditions:
\begin{listedense}
\item all inputs are orthogonal square matrices and/or unit column vectors;
\item the output matrix is a column vector.
\end{listedense}
\end{definition}

(We choose the term ``orthogonal-system-like'' because as we will show,
such a formula can be reorganized as a
product $M\cdot V$ of an orthogonal matrix with a column vector,
i.e. as the specification of an orthogonal system of linear equations.)

\begin{definition}
\label{def:SFT}
Let $K$ be a finitely generated semiring.
An instance of problem $\SFT{\K}$ (\emph{``sum-free formula partial trace''})
consists of an order $N \times 1$
OSL tensor formula $F$ over semiring $\K$ and a positive
integer $k$; the problem consists in
deciding whether the $\enth{k}$ partial trace of
$\left( \val{\K}^{N,1}(F) \right) \cdot
\Transpose{\val{\K}^{N,1}(F)}$
is greater than some predetermined constant
$\alpha$, $1/2 \le \alpha < 1$.
In the ``promise version'' of $\SFT{\K}$, no instance can yield a 
$\enth{k}$ partial trace which evaluates in the
interval  $[1-\alpha, \alpha]$.
\newline
We also define a ``nonzero version'' to $\SFT{\K}$, as the problem
which consists in deciding whether the $\enth{k}$ partial trace of
$\left( \val{\K}^{N,1}(F) \right) \cdot
\Transpose{\val{\K}^{N,1}(F)}$ is nonzero.
\end{definition} 

The following propositions show that
basic questions on inputs for problem $\SFT{\K}$ can be
answered in polynomial time.

\begin{proposition}
\label{prp:max-diameter}
\emph{\cite{beho01}}\
If~$F$ is a tensor formula of depth~$d$ which has input
matrices of diameter at most~$p$, then $|F|\leq p^{2^d}$, and there
exists a formula which outputs a matrix of exactly this
diameter. \emph{(Proof omitted.)}
\end{proposition}

\begin{proposition}
\label{prp:is-formula}
\emph{\cite{beho01}}\
Testing whether a string encodes a valid tensor formula and if so,
computing its order, is feasible in deterministic polynomial time.
\emph{(Proof omitted.)}
\end{proposition}

%


\section{From gate arrays to tensor formulas to gate arrays}
\label{sec:ctof:ftoc}

In this section we show how to encode the description of
a reversible or quantum gate array into a OSL tensor formula over
the appropriate semiring, and conversely, how to compute from an
OSL formula $F$ a gate array which will later used as a mean to solve
an SFT instance built from $F$.


\subsection{From arrays to formulas}
\label{subsec:ctof}

\begin{lemma}
\label{lem:array-goes-tensor-formula}
Let $C$ be a gate array operating on~$n$ wires,
whose gates can be described with orthogonal matrices
over semiring $\K$.
There is a logspace computable function which, given a suitable
coding of $C$, computes a tensor formula~$F(C)$ of logarithmic
depth such that for each
$x=(x_1,\ldots,x_n)\in\{0,1\}^n$,
$$C(x)=\val{\mathcal{\K}}^{n,1}(F(C)\cdot {d_x}),$$
where
$d_x=\bigotimes_{i=1}^n \chi_i$,
$\chi_i = e_2^1$ if $x_i=0$, and
$\chi_i = e_2^2$ otherwise.
\end{lemma}

\begin{proof}
Let~$C$ have $m$ levels and let~$C_i$ denote the $\enth{i}$
level, with~$C_1$ the left-most and~$C_m$ the right-most.
We describe how to construct an equivalent
tensor formula~$M(C)$ from~$C$ assuming
that~$\mathsf{0}$  and $\mathsf{1}$ are encoded by~$e_2^1$ and
$e_2^2$, \resp\
(for quantum arrays,
that~$\ket{0}$ and $\ket{1}$ are encoded by~$e_2^1$ and $e_2^2$, \resp).
We distinguish two cases.

(\emph{i})\ If each gate of $C_i$ acts on consecutive wires, that is, if 
$C_i$ contains $\ell \ge 1$ gates $H_1, \ldots, H_\ell$,
acting on wires $j_1$ to $k_1$, $\ldots$, $j_\ell$ to $k_\ell$,
with $j_1 \le k_1 < j_2 \cdots k_{\ell-1} < j_\ell \le k_\ell$,
then
$$M(C_i)=\left(
I_2^{\otimes j_1-1}\otimes H_1\otimes I_2^{\otimes j_2-k_1-1} \otimes \cdots
\otimes H_{\ell}\otimes I_2^{\otimes n-k_{\ell}} \right)$$
is the orthogonal matrix of order $2^n\times 2^n$ describing the
action of the $\enth{i}$ level of $C$. 

(\emph{ii})\ If~$C_i$ contains gates acting on nonadjacent wires, 
then choose a permutation $\sigma$ of the wires which brings
next to each other those wires which are involved in the same gate.
Denote by $D_i$
the $\enth{i}$ level reorganized in this way; its action on the
(permuted) wires is described with a formula $M(D_i)$ built as in
case (\emph{i}) above. The permutation is implemented by inserting
between levels $i-1$ and $i$ extra depth
levels consisting of swap gates, which are
collectively described by a formula $P_{\sigma}$;
it is undone with other extra levels, inserted between $i$ and $i+1$
and described by ${P}_{\sigma^{-1}}$.
Any permutation can be expressed as a product of a polynomial number
of cycles of the form
$(j, j+1,\ldots,k-1,k)$, with $j<k$; therefore it suffices to describe
the formulas $P_{j,k}(C)$
and $\bar{P}_{j,k}(C)$ which implement this cycle and its inverse, \resp.
$\bar{P}_{j,k}(C)$ which implements its inverse.
The reader can verify that\footnote{Note that according to the usual
convention, the input-to-output direction in a gate array is left-to-right,
while in its matrix representation, the array's action on its input
is given as a product of orthogonal matrices with a column vector,
and is read right-to-left.}  

$$ P_{j,k}(C) =\left( I_2^{\otimes j-1}\otimes T_{j,k}\otimes
I_2^{\otimes n-k} \right),
\quad \text{where}
\quad  
T_{j,k}=\prod_{i=1}^{k-j-1} \left(I_2^{\otimes k-j-i}\otimes
      T\otimes I_2^{\otimes i-1}\right),
$$
and
$$ \bar{P}_{j,k}(C) =\left( I_2^{\otimes j-1}\otimes \bar{T}_{j,k}\otimes
I_2^{\otimes n-k}\right),
\quad \text{where} \quad
\bar{T}_{j,k}=\prod_{i=1}^{k-j-1} \left(I_2^{\otimes i-1}\otimes
T\otimes I_2^{\otimes k-j-i} \right);$$
with $\sigma = ((j_1 \cdots k_1)\cdots (j_\ell \cdots k_\ell))^{-1} $,
this yields
$$P_{\sigma}(C) = \bar{P}_{j_1,k_1}(C) \cdots  \bar{P}_{j_\ell,k_\ell}(C)  
\quad \text{and} \quad
P_{\sigma^{-1}}(C) = {P}_{j_\ell,k_\ell}(C)  \cdots  {P}_{j_1,k_1}(C),$$
so that
$$M(C_i)= P_{\sigma^{-1}}(C) \cdot M(D_i) \cdot  {P}_{\sigma}(C).$$
A sample construction for $j=1$ and $k=4$ is depicted in
    Figure~\ref{fig:simulation}.
  \begin{figure}[tbp]
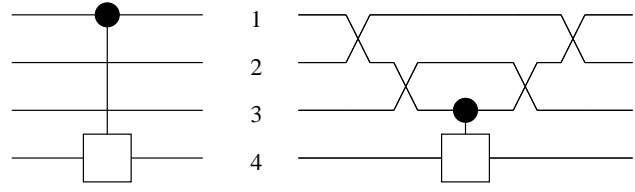

    \begin{center}
       \input simulation 
      \caption{Simulating an arbitrary controlled-not by a controlled-not
        acting on neighboring wires.}
      \label{fig:simulation}
    \end{center}
  \end{figure}

The complete tensor formula~$F(C)$ is given by
$$F(C) =\prod_{i=1}^m M(C_{i}),$$
which can be parenthesized in order to have logarithmic depth.
It is readily verified that for
each $x\in\{0,1\}^n$
$$C(x)=\val{\K }^{n,1}(F(C) \cdot {d_x}).$$
Formula~$F(C) $ is logspace constructible from $C$:
in particular, a permutation $\sigma$ suitable for
case (\emph{ii}) can be built by
choosing a reorganization $D_i$ of level $C_i$ in which
the gates $H_1, \ldots, H_\ell$,
act on wires $j_1$ to $k_1$, $\ldots$, $j_\ell$ to $k_\ell$,
such that $1=j_1$,  $ k_1 +1 = j_2$,  $ k_{\ell-1} +1 = j_\ell$;
then the cyclic decomposition of $\sigma$ 
has the form $(1,2,3,\ldots,h_1)(2,3,\ldots,h_2)\cdots$
where for each $i\ge 2$, the wires $1,2,\ldots,i-1$ are left untouched by
the $\enth{i}$ cycle.
\end{proof}


\subsection{From formulas to arrays} \label{subsec:ftoc}

In the formula-to-array part, one must deal with the fact that an
OSL formula may contain matrices of various sizes, and
column vectors at atypical locations. The latter may be regarded
a nonstandard or disorderly manner of specifying the array's inputs.
Matrices of nonstandard orders, however, cannot be readily 
interpreted in terms of Boolean or quantum computation: 
one may accept to work with many-valued bits and qubits,
or the matrices may be padded in order to turn their orders
into powers of $2$, which is the option we choose in this paper.

\begin{lemma}
\label{lem:embed-formula}
There exists a polynomial-time algorithm which turns an
OSL tensor formula $F$ over semiring $\K$
into a formula $\Pi(F)$ where all subformula sizes are
powers of $2$, and whose output is

\centerline{
$\left[ \begin{matrix} \val{\mathcal{\K}}^{n,1}(F) \\  0
\end{matrix}  \right], $}
where $0$ denotes a (possibly empty) null block.
\end{lemma}

\begin{proof}
For an integer $n \ge 0$, let $\pi(n)$ denote the smallest
power of $2$ greater than or equal to $n$. 
We also define a unary operator $\pi$ which acts as follows on a matrix $A$:
\begin{listedense}
\item if $A$ is a $n \times n$ square matrix, then $\pi(A)$
is  a $\pi(n) \times \pi(n)$ block-diagonal square matrix consisting in
a copy of $A$ at the top left position and a copy of the
identity matrix $I_{\pi(n)-n}$ at the bottom right;
\item if $A$ is a $n \times 1$ column vector, then $\pi(A)$
is  $\pi(n) \times 1$ with the entries of $A$ at the first
$n$ positions, and value $0$ in the $\pi(n)-n$ others;
\item if $A$ is neither of the above, then $\pi(A)$
is undefined.
\end{listedense}
Whenever $A\cdot B$, $\pi(A)$ and $\pi(B)$ are defined, we have
$\pi(A\cdot B)=\pi(A)\cdot \pi(B)$, so that in the simple case where $F$ does not
contain any occurrence of the Kronecker product, $\Pi(F)$ is built by
replacing each atomic subformula of $F$ with its image by $\pi$.
\newline
This does not work in general.
Consider for example the formula $(A \otimes B)\cdot(V \otimes W)$
where $A$ and $B$ are $33 \times 33$ and $35 \times 35$, respectively,
and 
$V$ and $W$ are $21 \times 1$ and $55 \times 1$, respectively:
the orders of $(\pi(A) \otimes \pi(B))$ and $(\pi(V) \otimes \pi(W))$
do not match. There also  exist cases where the orders match
but  the entries of
$(A \otimes B)\cdot(V \otimes W)$ are not consecutive
in the column vector
$(\pi(A) \otimes \pi(B))\cdot(\pi(V) \otimes \pi(W))$.
Some subformulas may even yield matrices
which are neither square nor column vectors.
\newline
Nevertheless, we claim that if matrices $\Pi(A)$ and $\Pi(B)$ are available,
then there exists permutations $Q$ and $Q'$ and a block $H$ such that

$$
Q \cdot (\Pi(A) \otimes \Pi(B)) \cdot Q' =
\left[ \begin{array}{cc} A\otimes B & 0 \\ 0 & H \end{array} \right] ,
$$

where $Q$ and $Q'$ can be specified with
polynomial-size sum-free tensor formulas.
(Note that $H$ is orthogonal whenever both $A$ and $B$ are.)
In the special case where both $A$ and $B$ are column vectors,
$Q'=I_1$ and the claim reads

\centerline{
$Q \cdot (\Pi(A) \otimes \Pi(B)) =
{\left[ \begin{matrix} A\otimes B\\ 0 \end{matrix} \right]}$.}

We first show how to reorder the lines of $\Pi(A) \otimes \Pi(B)$
where both $A$ and $B$ are column vectors.  With
$\ A = \transpose{[\ x_1\ \cdots x_m\ ]}\ $   
and 
$\ B = \transpose{[\ y_1\ \cdots y_n\ ]},\ $
let $\mu = 2^j \ge \pi(m) $, $\sigma = \mu-m$, $\nu = 2^k \ge \pi(n) $,
and $\tau = \nu-n$.  We start with
$$
\Pi(A) = \transpose{[\ x_1\ \cdots x_m\ \bar{x}_{m+1} \cdots \bar{x}_{\mu}\ ]},
\quad \quad
\Pi(B) = \transpose{[\ y_1\ \cdots y_n\ \bar{y}_{n+1} \cdots \bar{y}_{\nu}\ ]}
$$
$$
\text{and} \quad \Pi(A) \otimes \Pi(B) =    
\transpose{[\ x_1y_1\ \ x_1y_2\ \cdots  x_1\bar{y}_{\nu}\ \
x_2y_1\ \ x_2y_2\ \cdots \bar{x}_{\mu}\bar{y}_{\nu}\ ]} ;
$$
the $\bar{x}_i$'s and $\bar{y}_i$'s are the elements added by padding.
Multiplying to the left with the stride permutation $P_{\nu}^{\mu\nu}$ gives
$$
P_{\nu}^{\mu\nu} \cdot \left(\Pi(A) \otimes \Pi(B)\right) =    
\transpose{[\ x_1y_1\ \ x_2y_1\ \cdots  x_{\mu} \bar{y}_1\ \ x_1y_2\ \ x_2y_2\
\cdots \bar{x}_{\mu}\bar{y}_{\nu}\ ]} .
$$
Next we multiply with the matrix 
$$
R_{n}^{\mu\nu} =
\left[ \begin{array}{cc} 
P_{\mu}^{n\mu} & 0 \\ 0 & \left( N^{\mu\tau} \right)^k 
\end{array} \right] 
$$
where $ N^{\mu\tau} = I_\tau \otimes P_2^{\mu}$.
The reader can verify that 
$$
R_{n}^{\mu\nu} \cdot  P_{\nu}^{\mu\nu} \cdot \left(\Pi(A) \otimes \Pi(B)\right) =  
\transpose{[\ x_1y_1\ x_1y_2\ \cdots  x_1y_n\ \cdots x_my_n\ H\ ]} =
\transpose{[\ (A \otimes B)\ H\ ]}
$$
where $H$ is a size $\mu\nu-mn$ block whose first $n\sigma$ entries are
$\bar{x}_{m+1}y_1, \ldots, \bar{x}_{\mu}y_n$ and the 
other positions contain a permutation of 
${x}_{1}\bar{y}_{n+1}, \ldots, {x}_{1}\bar{y}_{\nu}, \ldots,
\bar{x}_{\mu}\bar{y}_{\nu}.$
\newline
There remains to show how to build matrices 
$P_{\nu}^{\mu\nu} $ and $R_{n}^{\mu\nu} $
with polynomial-size sum-free tensor formulas.
By Proposition \ref{prp:basic-stride},
it is readily verified that 
$P_{\nu}^{\mu\nu} = \left(P_{2}^{\mu\nu}\right)^k $, and 
that for any $\ell \ge 1$, the induction formula
$P_{2}^{2^{\ell +2}} =
\left(P_{2}^{2^{\ell +1}} \otimes I_2 \right) \cdot
\left( I_{2^\ell } \otimes P_{2}^{4}\right)$ 
yields for the matrix 
$P_{2}^{\mu\nu}$ a quadratic-size tensor formula with 
input nodes for $I_2$ and $P_24$.
Meanwhile, $R_{n}^{\mu\nu} = \left(S_{n}^{\mu\nu}\right)^k$, where 
$$
S_{n}^{\mu\nu} =
\left[ \begin{array}{cc} 
P_{2}^{n\mu} & 0 \\ 0 & N^{\mu\tau} \end{array} \right] .
$$
In order to build this matrix, let
$$
  U =
\left[ \begin{array}{cc} 
P_{2}^{2n} & 0 \\ 0 & I_{2\tau} \end{array} \right] 
$$
and 
$P_{2}^{n\mu} =
\left(P_{2}^{2n} \otimes I_{2^{j-1}} \right) \cdot
\left( I_{n} \otimes P_{2}^{\mu}\right)$ by Proposition
\ref{prp:basic-stride};
observe that 
$$
\left( U \otimes I_{2^{j-1}} \right) \cdot
\left( I_{\nu} \otimes P_{2}^{\mu} \right)
=
\left[ \begin{array}{cc} 
P_{2}^{2n} \otimes I_{2^{j-1}} & 0 \\ 0 & I_{\tau\mu} \end{array} \right] 
\cdot
\left[ \begin{array}{cc} 
I_n \otimes P_{2}^{\mu} & 0 \\ 0 & I_{\tau} \otimes P_{2}^{\mu} \end{array} \right] 
=
\left[ \begin{array}{cc} 
P_{2}^{n\mu} & 0 \\ 0 & I_{\tau} \otimes P_{2}^{\mu} \end{array} \right] =
S_{n}^{\mu\nu}.
$$
Expressed in this way, matrix $R_{n}^{\mu\nu}$ can be built with
a polynomial-size sum-free tensor formula, where
matrix $U$ is either
given explicitly by a made-to-purpose gate if $n$ is the diameter of
an input matrix, or built inductively in the
case where $n = \pi(p)$ for some $p$,
because in this case $U = P^{2n}_2$.
\newline
The same technique applies to reorder the lines
for arbitrary matrices $A$ and $B$; in this
case the $x_i$'s and $y_i$'s are lines and each $x_iy_j$ in the
above equations must
be read as $x_i \otimes y_j$. The claim for the existence of a matrix
$Q'$ which reorders the columns is proved in a dual manner.
\newline
Let $F$ be an OSL formula; 
the following algorithm builds a formula $\Pi(F)$ which satisfies
the conditions of the Lemma, by recursively defining $\Pi(G)$ for
each subformula $G$ of $F$.
\begin{listedense}
\item For each atomic subformula  $G$, let $\Pi(G) = \pi(G)$.
\item Repeat recursively from the leaves toward the root of $F$:
for each subformula $G = H \circ K$ for which $\Pi(H)$ and $\Pi(K)$
have already been computed and $\circ \in \{\cdot,\otimes\}$: 
\begin{listedense}
\item
if $\circ$ is ``$\otimes$'' then 
let $\ \Pi(G) = Q \cdot (\Pi(H) \otimes \Pi(K)) \cdot Q'\ $ 
and insert the appropriate subformulas for $Q$ and $Q'$
(note that $\Pi(Q)=Q$ and $\Pi(Q')=Q'$);
\item
otherwise $\circ$ is ``$\cdot$'': if the orders of 
$\Pi(H)$ and $ \Pi(K)$ match, then 
let $\ \Pi(G) = \Pi(H) \cdot \Pi(K); $
else they differ by a power of $2$ and the smaller matrix
must undergo some padding, that is, either
$\Pi(G) = (I_2^{\otimes i} \otimes \Pi(H)) \cdot \Pi(K)$, or
$\Pi(G) = \Pi(H) \cdot ((e_2^1)^{\otimes i} \otimes \Pi(K)),$
for an appropriate $i$.
\end{listedense}
\end{listedense}
\end{proof}

\begin{lemma}
\label{lem:tensor-formula-goes-array}
There is a polytime computable function which, from
a OSL tensor formula $F$ over semiring $\K$,
computes a polynomial-size gate array~$C(F)$ 
whose input is represented with
a unit vector $V$, whose action over the inputs is
given by an orthogonal matrix $M$, and such that 
matrices $MV$ and 
$\ \val{\mathcal{\K}}^{n,1}(F)\ $
satisfy

\centerline{
$ MV = 
\left[ \begin{matrix}
\val{\mathcal{\K}}^{n,1}(F)\ \\ 0 \end{matrix} \right] $,
}
where $0$ denotes a (possibly empty) null block.
\end{lemma}

\begin{proof}
The formula $\Pi(F)$ is used as a specification for a gate array $C(F)$.
For each atomic subformula $G$ of $F$, either $G$ is $m \times m$
for some $m \le |F|$, where $|F|$ is the diameter of $F$,
and $\Pi(G)$ is interpreted as the specification of a gate with
$\log_2 \pi(m) = \lceil \log_2 m \rceil$ inputs,
or $G$ is $m \times 1$ and $\Pi(G)$ 
specifies the probability amplitudes for all possible
combinations of values of
$\log_2 \pi(m) = \lceil \log_2 m \rceil$ input bits or qubits.
In the former case, a polynomial-size 
array of elementary gates implements the  
operation specified by $\Pi(G)$; in the latter case,
a size $m^{O(1)}$ array is built to take as input
some constant unit vector (say $e^1_{\Pi(m)}$)
and yield as output the vector $\Pi(G)$.
Next, working recursively from the leaves toward the root of $\Pi(F)$,
the interior nodes are interpreted as 
specifications for combining the subarrays either in a
sequential (nodes labelled ``$\cdot$'') or 
parallel (nodes labelled ``$\otimes$'') manner.
The resulting gate array has polynomial size and
satisfies the conditions of the lemma.
\end{proof}


\section{Complexity results }
\label{sec:complex}

Over the Boolean semiring, a column vector is a unit vector
as soon as it is nonzero,
so that the standard, promise and nonzero versions of problem SFT coincide.

\begin{theorem} \label{thm:boolean}
Over the Boolean semiring, problem SFT is $\Poly$-complete under
logspace reducibility.
\end{theorem}

\begin{proof}
Given a size $n$ instance $(F,k)$ of $\SFT{\B}$, 
we use Lemma
\ref{lem:tensor-formula-goes-array}\
to build an equivalent reversible gate array $C(F)$
over $N=n^{O(1)}$ bits, and we compute the output value of
each of these bits (i.e. we solve $N$ instances of the usual
Boolean circuit value problem). This yields a
combination of $N$ values which corresponds to a given position
along the diagonal of 

\centerline{
$\left( \val{\mathcal{\B}}^{2^N,1}(F) \right) \cdot 
\Transpose{\val{\mathcal{\B}}^{2^N,1}(F)},$}

under the convention that combinations $00\cdots 0$, $\ldots$,
$11\cdots 1$ correspond to lines (and columns) $1,\ldots,2^N$, respectively. 
The hardness part consists in using Lemma
\ref{lem:array-goes-tensor-formula}\ to reduce
the $\Poly$-complete circuit value problem \cite{lad75}\
to an instance of  $\SFT{\B}$.
\end{proof}

\hspace{0.1in}

For the quantum and probabilistic cases we are mainly interested in
the promise version of SFT, which gives us a striking description for
the difference between complexity classes \BPP\ and  \BQP.

\begin{theorem} \label{thm:QPPandBQP}
The promise version of problem SFT($\Q$) is complete for the 
class $\BQP$, under logspace reducibility.
\end{theorem}

\begin{proof}
The hardness part is a generic reduction.
Using Definition \ref{def:PPandBPP}, we start with
a $m$-leveled gate array $C$ on $n$ qubits numbered $1$ to
$n$ whose accepting subspace is defined by setting qubit $1$ to $\ket{1}$,
and whose gates are defined with unitary matrices over $\Q$.
Denote by $f_C$ the probability that qubit $1$ be
projected to $\ket{1}$ when the measurement takes place.
We use Lemma \ref{lem:array-goes-tensor-formula}\ 
to build from $C$ an equivalent tensor formula 
$F(C) =\prod_{i=1}^m M(C_{i}).$
Meanwhile we define for the array's input qubits a
tensor product $V$ of $n$ unit vectors of size $2\times 1$.
An easy induction on $j$ shows that 

\centerline{
$\val{\mathcal{\Q}}^{2^n,1} \left( \prod_{i=1}^j M(C_{i}) \cdot V \right)$}

is exactly the vector of amplitudes after level $j$ in $C$. 
Thus the last $2^{n-1}$ entries along the diagonal of

\centerline{ $\left( \val{\mathcal{\Q}}^{2^n,1}( F(C)\cdot V) \right)
\cdot \Transpose{\val{\mathcal{\Q}}^{2^n,1}(F(C)\cdot V)} $}

add up to the value of $f_C$, and
the original array's input is accepted iff
this partial trace exceeds the threshold 
by which acceptance by $C$ was defined.
Scrutiny of the reduction shows that the constraint 
on $f_C$ is transported
intact from the description of $C$ to the $\SFT{\Q}$ instance $F(C)\cdot V$.
\newline
In the other direction, we use
Lemma \ref{lem:tensor-formula-goes-array}\ 
to translate an instance $(F,k)$ for $\SFT{\Q}$ into the
description of a quantum gate array over $m$ qubits,
$m \ge \log_2 n$, and of its inputs;
the $\enth{k}$ partial trace of

\centerline{
$\left( \val{\mathcal{\Q}}^{2^m,1}(F) \right) \cdot
\Transpose{\val{\mathcal{\Q}}^{2^m,1}(F) }$}

represents the probability that the output qubits of this
array be projected onto the direct sum of the dimension-$1$
subspaces generated by
$\ket{2^m-1}=\ket{1\cdots 11}$,
$\ket{2^m-2}=\ket{1\cdots 10}$,
$\ket{2^m-3}=\ket{1\cdots 01}$,..., and $\ket{2^m-k}$.
The promise on the partial trace is transported
unmodified from the input tensor formula to the quantum gate array.
\end{proof}

The argument described above can be used to prove that
the ``standard'' (non-promise) version of problem SFT($\Q$) is complete for 
\PP, defined by removing the constraint from definition \ref{def:BQP}.
Finally, when the proof is applied to the ``nonzero'' version of
problem SFT($\Q$), a completeness statement is obtained  for the class \NQP.

\hspace{0.1in}

Finally, we consider problem SFT over
the semiring of the nonnegative rational numbers.
Note that, just as in the quantum case, the entries in
the column vectors are regarded as probability \emph{amplitudes}.
All the gates do in a classical reversible array is
permute the different vector components without ever mixing or combining them;
no interference ever takes place and
it does not matter in terms of the final result, whether the probabilities
are represented as such or as amplitudes.

\begin{theorem} \label{thm:PPandBPP}
Problem SFT($\Q^+$) is \PP-complete under logspace reducibility.
\end{theorem}

\begin{proof}
For a generic reduction, we start with a reversible gate array $C$
whose input is a string of $N=s(n)+t(n)$ bits,
where the initial $s(n)$ bits are the ancillary bits, 
all set to $0$, and the
other $t(n)$ bits are random.
By Lemma \ref{lem:array-goes-tensor-formula}, $C$ and its input can be
encoded into $F(C) \cdot V$, where
the $2^N \times 1$ unit vector $V$ specifies the inputs,
i.e. a bit string $ c_1 \cdots c_{s(n)} d_1 \cdots d_{t(n)} $
which satisfies the conditions 
\begin{latindense}
\item $c_i = 0$ for all $i \le i \le s(n)$,    and
\item all combinations of values for the random bits $d_1 \cdots d_{t(n)}$
are equally likely.
\end{latindense}
The corresponding $2^{t(n)}$ entries in the vector 
$\val{\mathcal{\Q^+}}^{2^N,1}(V) $
carry value $1/\sqrt{2^{t(n)}}$; all others contain $0$.
We demand wlog that $t(n)$ be even; dealing with the random bits pairwise
enables us to ensure that no irrational values are necessary.  Then 

$$
V\ =\ (e_1^1)^{\otimes s(n)} \otimes
{\left[ \begin{array}{c}
1/\sqrt{2} \\ 1/\sqrt{2} 
\end{array} \right]}^{\otimes t(n)}\ =\
(e_1^1)^{\otimes s(n)} \otimes
{\left[ \begin{array}{c}
 1/2 \\ 1/2 \\ 1/2 \\ 1/2 
\end{array} \right]}^{\otimes \frac{t(n)}{2}} .
$$

Let the acceptance condition be that bit $c_1$ has value $1$ at the
output level. This corresponds to the first
$2^{N-1}$ positions along  the diagonal of 
$\left( \val{\mathcal{\Q^+}}^{2^N,1}(F(C)\cdot V) \right) \cdot
\Transpose{\val{\mathcal{\Q^+}}^{2^N,1}(F(C)\cdot V)} $.

In the other direction,
consider an instance $(F,k)$ for $\SFT{\Q^+}$.
We have discussed in Section \ref{subsec:ftoc}\  how the
column vectors and square matrices are interpreted as ``inputs''
and ``gates'' in the equivalent array, through the construction of
a formula $\Pi(F)$ where all matrices have 
orders which are powers of 2. 
We add extra steps to the construction of $\Pi(F)$ in order to
enforce the further condition, that all fractions have a power of
$2$ as denominator. 
\newline
Consider a $n \times 1$ unit
vector $v_i = \transpose{[ \frac{a_1}{d} \cdots  \frac{a_n}{d} ]}$,
where $a_1^2 + \cdots +a_n^2 = d^2$.
Let $d$ not be a power of $2$: $d < \pi(d)$.
The reader can verify that there exist integers 
$b_1,\ldots,b_p$  such that 
$\pi(d)^2 = a_1^2 + \cdots +a_n^2 + b_1^2 + \cdots + b_p^2\  $
and $p \le 3 \lceil \log_2 d \rceil $.
Let $q = \min\{ 2^{2j} : 2^{2j} >  n + 3 \lceil \log_2 d  \rceil \}$,
and embed $v$ into the $q \times 1$ vector 

$$
\transpose{ \left[ \
\frac{a_1}{\pi(d)}\ \cdots\ \frac{a_n}{\pi(d)}\ 0\ \cdots\ 0\ 
\frac{b_1}{\pi(d)}\ \cdots\ \frac{b_p}{\pi(d)}\ \right] } ,
$$

which can be interpreted as a distribution of probability amplitudes for 
$\log_2 q$ input bits.
Denote by $\delta_i$ the fraction $d/\pi(d)$.
Repeating this process on each input column vector yields an instance
($G,k$) where the resulting partial trace is the same one obtained from
$(F,k)$, times a factor $\Delta^2 = \prod_i \delta_i^2$.
If we accept instance 
$(F,k)$ whenever the partial trace is above a threshold $\alpha$,
then there exists
a probabilistic polytime Turing machine $M$ which accepts
$(G,k)$ with probability above $\frac{\alpha}{\Delta^2}$.
\newline
The algorithm of $M$ 
is divided into three phases; the first consists in building the
new instance $(G,k)$ from the original $(F,k)$, the second in 
choosing nondeterministically a column vector to give as input to the
equivalent array $C(G)$, and the third in deterministically simulating
$C(G)$ on its input. In the second step $M$ nondeterministically selects
values for the bits in the string $d_1\cdots d_{t(n)}$;
the preprocessing step has organized their probability distribution
in order to ensure that this can be done with
a sequence of nondeterministic binary choices,
followed by a look-up into a table which is linear in size
and is computed from the column vectors in $(F,k)$. 
\end{proof}

The reader can verify that this proof can be rewritten in terms
of the promise problem SFTP($\Q^+$) and the complexity class \BPP;
in the second part of the proof the cutpoint and the size
of the empty interval can be modified, however.
Meanwhile, the complexity of the nonzero version is obtained
with a straightforward
application of the above argument.

\begin{corollary}
\label{cor:BPP}
The promise and nonzero versions
of problem SFT($\Q^+$) are \BPP-complete and \NP-complete,
respectively, under logspace reducibility.
\end{corollary}

\begin{figure}

\begin{center}
\begin{tabular}{|c|c|c|c|}
\hline
Semiring/Version & Standard & Promise & Nonzero \\
\hline
\hline
$(\Q,+,\cdot)$ & \PP & \BQP\ & \NQP \\
\hline
$(\Q^+,+,\cdot)$ & \PP\ & \BPP\ & NP\ \\
\hline
$(\bool,\vee,\wedge)$ & \multicolumn{3}{|c|}{ \Poly } \\
\hline
\end{tabular}

\caption{Summary of completeness results}
\end{center}

\end{figure}


\section{Conclusion}
\label{sec:conclusion}

Through the study of problem SFT, we have developed a common
algebraic description for polynomial-time complexity classes,
where the choice of the semiring determines the complexity class.
For the inclusion chain $\Poly \subseteq \BPP \subseteq \BQP$, in particular,
the classical model of polytime probabilistic computation turns 
out to be a special case of polytime quantum computation  where
interference between computations is ruled out.\\

\textbf{Acknowledgements}

We wish to thank Lance Fortnow and John Watrous for helpful discussion
and pointers to useful references. Insightful comments by Gilles Brassard,
Michele Mosca and Pierre McKenzie were also appreciated.

\setlength {\baselineskip} {12pt}

\bibliographystyle{plain} 
\bibliography{komp}


\end{document}